\documentclass[prd,nofootinbib,showpacs,twocolumn]{revtex4}
\usepackage{graphicx}
\usepackage{bm}

\begin{document}

\title{ Large-scale power in the CMB and new physics:
an analysis using Bayesian model comparison}
\author{Anastasia Niarchou, Andrew H. Jaffe and Levon Pogosian}
\affiliation{Blackett Laboratory, Imperial College London, SW7 2AZ, United Kingdom}

\date{\today}\medskip

\begin{abstract}
  One of the most tantalizing results from the WMAP experiment is the
  suggestion that the power at large scales is anomalously low when
  compared to the prediction of the ``standard'' $\Lambda$CDM model. The
  same anomaly, although with somewhat larger uncertainty, was also
  previously noted in the COBE data. In this work we discuss possible
  alternate models that give better fits on large scales and apply a
  model-comparison technique to select amongst them. We find that models
  with a cut off in the power spectrum at large scales are indeed
  preferred by data, but only by a factor of 3.6, at most, in the
  likelihood ratio, corresponding to about ``$1.6\sigma$'' if
  interpreted in the traditional manner. Using the same technique, we
  have also examined the possibility of a systematic error in the
  measurement \emph{or prediction} of the large-scale power. Ignoring
  other evidence that the large-scale modes are properly measured and
  predicted, we find this possibility somewhat more likely, with roughly
  a $2.5\sigma$ evidence.
\end{abstract}

\pacs{98.80.Cq}


\maketitle

\section{Introduction}
\label{sec:intro}

\begin{figure}[tbp]
  \centering\includegraphics[width=0.67\columnwidth,angle=270]{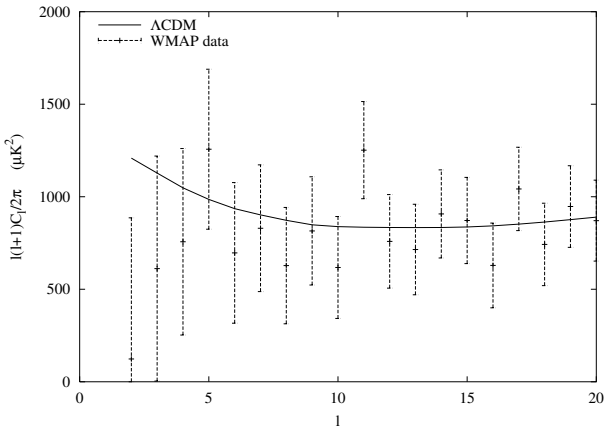}
  \caption{The CMB power spectrum at low $\ell$ as measured by WMAP. The
    solid line is the best fit using the ``standard'' power law
    $\Lambda$CDM model. Note that the error bars at low multipoles are
    almost entirely due to cosmic variance.}
  \label{fig:Cl}
\end{figure}

The recent WMAP results \cite{BennettWMAPbasic03} have provided a spectacular
view of the early Universe. One of the most intriguing results offered
by the WMAP team is that the CMB anisotropy power on the largest angular
scales seems to be anomalously low \cite{BennettWMAPbasic03,WMAP03Spergel}. In
fact, the WMAP team report that this result has a high statistical
significance, quoting a probability ranging from just under 1\% to
$2\times10^{-3}$ for such a result, depending on the details of the
analysis. This low power can be seen in two complementary ways. First,
in the CMB power spectrum, $C_\ell$, the quadrupole ($\ell=2$) and
octopole ($\ell=3$) both seem low in comparison to the smooth ``best
fit'' model, as shown in Figure~\ref{fig:Cl}.
The latter is selected from the array of models with a flat geometry and
nearly scale invariant, adiabatic primordial fluctuations.

\begin{figure}[tbp]
  \centering\includegraphics[width=0.67\columnwidth,angle=270]{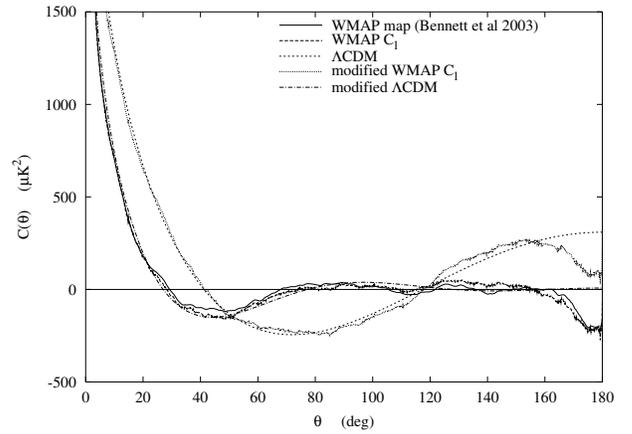}
  \caption{The correlation function, $C(\theta)$ as computed from the
    WMAP team, from the pixelized map (solid line); using the $C_\ell$s
    measured by WMAP (long dashed line), using WMAP's best fit $C_\ell$
    (short dashed), using the WMAP data with $C_2$ and $C_3$ changed to
    equal those of the best fit (dotted), and using the best fit
    $C_\ell$s with lowered values of $C_2$ and $C_3$ (dot-dash).}
  \label{fig:ctheta}
\end{figure}

The low power seems particularly striking when the CMB anisotropy correlation
function,
\begin{equation}
\label{eq:Ctheta}
  C(\theta)\equiv\langle T({\bf\hat n}) T({\bf\hat m}) \rangle
  \qquad\textrm{with}~ {\bf\hat n}\cdot{\bf\hat m}=\cos\theta
\end{equation}
is examined: it is very near zero for $\theta\gtrsim60^\circ$. Note that

the average implied by the angle brackets has several different,
inequivalent, interpretations: The WMAP team estimate the correlation
function calculated as the simple average over pixels at a given
separation. If we interpret the average as an ensemble average, however,
we can relate the correlation function to the power spectrum, $C_\ell$:
\begin{equation}
  \label{eq:ClCtheta}
  C(\theta) = \sum_\ell \frac{2\ell+1}{4\pi} C_\ell P_\ell(\cos\theta)
\end{equation}
For a Gaussian distribution with enough samples, these two
definitions are nearly equivalent, since the pixel average will
approximate the ensemble average. We were able to reproduce the
character of the correlation function from the published angular power
spectrum, by summing the Legendre series in Eq.~(\ref{eq:ClCtheta}). In
fact, we obtained almost the same result by using the smooth best-fit
spectrum, but with the quadrupole and octopole lowered to the observed
levels, as also shown in Figure \ref{fig:ctheta}.  (In fact, the
correlation function in this case is actually flatter at
$\theta\sim180^\circ$ than those computed from the actual data: the
power in any of the correlation functions calculated from real data
shows a lower correlation amplitude than those calculated from smooth
power spectra.)  Conversely, raising the quadrupole and octopole in the
observed spectrum to the predicted levels removes the anomaly. This
exercise implies that the low power is just that: low \emph{power} at
low $\ell$, and due neither to a conspiracy of particular $C_\ell$
values nor to any non-Gaussian distribution of the multipole moments
themselves. Moreover, the apparently striking difference between the
measured and predicted $C(\theta)$ is due entirely to the low values of
the quadrupole and octopole. In this paper, we investigate the
statistical significance of these measurements.

In the following, we introduce the Bayesian model comparison
method in Section~\ref{sec:ModelComparison}, discuss models with
low primordial power in Section~\ref{sec:lowpower}, and a model of
experimental or theoretical systematic errors in
Section~\ref{sec:systematics}. We conclude with a discussion in
Section~\ref{sec:discussion}.

\section{Model Comparison}
\label{sec:ModelComparison}

The question remains, then: how significant is this observed low power?
Here, we shall answer this question using the technique of Bayesian
model comparison \cite{Jaynes2003,Loredo90}. This technique has been
used before in various cosmological contexts
\cite{JaffeOdds96,HobsonBridleLahav2002,SainiDarkEnergyEvidence03,JohnNarlikar}.

We start, as usual, with Bayes' theorem, which gives the posterior
probability of some theoretical parameters $\theta$ given data $D$
under the hypothesis of some model $m$:
\begin{equation}
  \label{eq:bayes}
  P(\theta | D I_m) = P(\theta | I_m) \frac{P(D|\theta I_m)}{P(D|I_m)}
\end{equation}
where $P(A|B)$ gives the probability or probability density of a
proposition $A$ given a proposition $B$ and, here, \emph{all}
probabilities are conditional, at least on the background information
$I_m$, which refers to the background information for a specific model
$m$. The model parameters $\theta$ (the list of which may actually
depend on which model $m$ we consider), have prior probability
$P(\theta|I_m)$. The likelihood function is $P(D|\theta I_m)$, and the
so-called ``evidence'' is
\begin{equation}
\label{eq:evidence}
P(D|I_m)=\int d\theta P(\theta|I_m) P(D|\theta I_m)
\end{equation}
which enforces the normalization condition for the posterior but is also
quite properly the probability of the data given model $m$, the ``model
likelihood''.

We can further factor the evidence as
\begin{equation}
  \label{eq:Ockham}
  P(D|I_m) = {\cal L}_m(\theta_{\rm max}) O_m
\end{equation}
where $\theta_{\rm max}$ are the parameters that maximize the likelihood
for model $m$, ${\cal L}_m(\theta) = P(D|\theta I_m)$, and $O_m$ is the
so-called ``Ockham Factor'' \cite{Jaynes2003}. The Ockham factor is
essentially the ratio of the prior probability volume to the posterior
probability volume. (This is most easily seen for the case where both
prior and posterior are uniform distributions. When both are Gaussian
distributions, the Ockham factor is the ratio of the determinants of the
covariance matrices, which is indeed the ratio of the $1\sigma$ volumes.)

In order to select among models, one usually employs the ratio of their
probabilities:
\begin{equation}
\label{eq:ratio}
\frac{P(m|DI)}{P(n|DI)}=\frac{P(m|I)}{P(n|I)}\frac{P(D|I_m)}{P(D|I_n)}=\frac{P(m|I)}{P(n|I)}B_{mn}
\end{equation}
Any experimental information is contained in the ratio of the evidence,
$B_{mn}$, which is referred to as the ``Bayes factor''. Lacking any
prior information preferring one model over the other,
Eq.~(\ref{eq:ratio}) only depends on the Bayes factor.
Eqns.~(\ref{eq:evidence}--\ref{eq:ratio}) imply that the Bayes factor incorporates the
essence of the Ockham razor: since the evidence is an average of the
likelihood function with respect to the prior on the parameters, simpler
models having a more compact parameter space are favored, unless more
complicated models fit the data significantly better. Bayes factors are
likelihood ratios, and can be interpreted roughly as follows, as
suggested in Ref.~\cite{DrellLoredoWasserman}: If \begin{math}
  1<B_{mn}\lesssim3 \end{math}, there is an evidence in favor of model
$m$ when compared with $n$, but it is almost insignificant. If
\begin{math}3\lesssim B_{mn}\lesssim20\end{math}, the evidence for $m$
is definite, but not strong. Finally, if
\begin{math}20\lesssim B_{mn}\lesssim150\end{math}, this evidence is
strong and for \begin{math}B_{mn}\gtrsim150\end{math} it is very strong.

We can also interpret the likelihood ratio in the same manner as we
compute the ``number of sigma'' by which a value or hypothesis is
favored. In this case the model is favored by $\nu\;\sigma$ with
$\nu=\sqrt{2\ln \left|B_{mn}\right|}$. Another useful interpretation, perhaps more
familiar to the engineering community, would be to use decibels, $0.1
\log_{10}{B_{mn}}$ \cite{Jaynes2003}.

The model comparison formalism outlined here \emph{requires} us to
specify alternatives to the ``fiducial'' standard model. Thus a sharper
version of our question might be: is it more probable that the data do
reflect a standard Big Bang, with nearly-scale invariant, adiabatic,
isotropic, Gaussian fluctuations, or do they come from a Universe with,
say, a cutoff in the power spectrum? Or could there be a problem in the
data analysis so that, say, the error bars are larger than thought, or
the reported results somehow exhibit an over-subtraction of large-scale
power? In the following we shall examine these possibilities.

The ``fiducial'' standard model is the best-fit model from
\cite{WMAP03Spergel}. It is a flat $\Lambda$CDM
Friedmann-Robertson-Walker universe, with baryon density
$\Omega_b=0.046$ and ``dark energy'' density $\Omega_\Lambda=0.73$ (in
units of the FRW critical density). It has a power-law initial matter
power spectrum with spectral index $n_s=0.99$ and a present-day
expansion rate of $H_0=100h\;\mbox{km}\;\mbox{sec}^{-1}\;\mbox{Mpc}^{-1}$
with $h=0.72$. The power spectrum amplitude is $A_s=0.855$, as defined
in the CMBFast program \cite{CMBFAST} and as used by the WMAP team
\cite{WMAP_Peiris}, related to the amplitude of fluctuations at $k_0=0.05\;\mbox{Mpc}^{-1}$.

The evidence for this model is simply the likelihood $P(D|\theta I_{\rm
  fiducial})$ evaluated at the best fit values of the parameters. We
calculate the likelihood using the code provided by the WMAP team
(\cite{WMAP_Peiris}), which correctly accounts for correlations between
values of $\ell$ and the non-Gaussian shape of the distribution.  For
the fiducial model it is equal to $0.00094$, which is the value that we
will need when comparing to other models.

\section{Low-power models}
\label{sec:lowpower}

\subsection{A flat Universe with a cutoff in the primordial spectrum}
\label{sec:cutoff}

The most obvious way to lower the CMB power spectrum is to lower
the power in the primordial density power spectrum $P(k)$
\cite{Yokoyama99,Bridle03,cline03,Feng03,ContaldiCutoff03}. Since
the CMB is the product of small fluctuations in the primordial
plasma, we can use linear theory. To each multipole $\ell$ there
corresponds a transfer function $T_\ell(k)$, such that
$\ell(\ell+1)C_\ell=2\pi\int d\ln{k} \; T_\ell(k) k^3 P(k)$. The
transfer function depends on the cosmological parameters, but is
peaked at approximately $k\eta_0\sim \ell$, where $\eta_0$ is the
current size of the universe, of order $\eta_0 \sim 1.5
\times10^{4}\;\mbox{Mpc}$. Lowering power at $k\lesssim
6\times10^{-4}\;\mbox{Mpc}^{-1}$ thus lowers the CMB power
spectrum for $\ell\lesssim4$.

A simple model where such a cutoff was imposed by
fiat was considered by Contaldi et al \cite{ContaldiCutoff03}.
They used the following form for the primordial spectrum:
\begin{equation}
  P(k)=P_0(k) \ \left[1-e^{-(k/k_c)^\alpha}\right] \ ,
\label{cutoffpk}
\end{equation}
where $P_0(k)=A k^n$ is the usual power law primordial spectrum.  They
rightly determine that the data favor a cutoff at $k_c\simeq (5{\rm-}6)
\times 10^{-4}\;\mbox{Mpc}^{-1}$. In \cite{ContaldiCutoff03} Contaldi et
al considered another class of models with the cutoff produced by
altering the shape of the inflaton potential.  Here, we concentrate on
the lower multipoles alone and consider the effect of varying
\emph{only} the location of the power cutoff using Eq.~(\ref{cutoffpk})
with $\alpha=1.8$. This reasonably assumes that there is enough freedom
in the model space to allow the high-$\ell$ spectra to adjust to fit the
data, and that the transfer function, $T_\ell(k)$, does not change much
at low $\ell$.

\begin{figure}[tbp]
  \centering\includegraphics[width=0.67\columnwidth,angle=270]{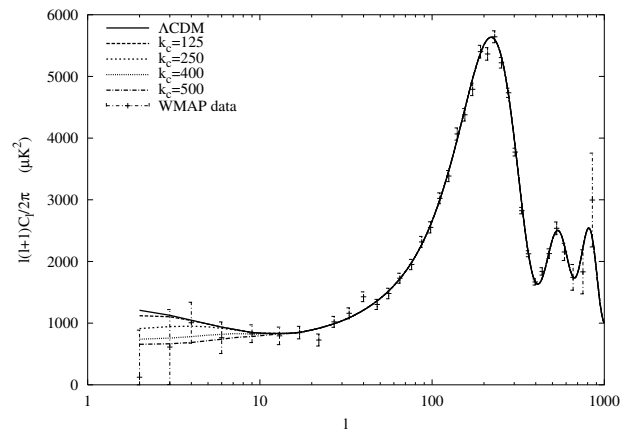}
  \caption{CMB power spectra for various values of the cutoff parameter
    $k_c$ of Eq.~\ref{cutoffpk}, measured in units of
    $10^{-6}\;\mbox{Mpc}^{-1}$.}
  \label{fig:Clcutoff}
\end{figure}
\begin{figure}[tbp]
  \centering\includegraphics[width=0.67\columnwidth,angle=270]{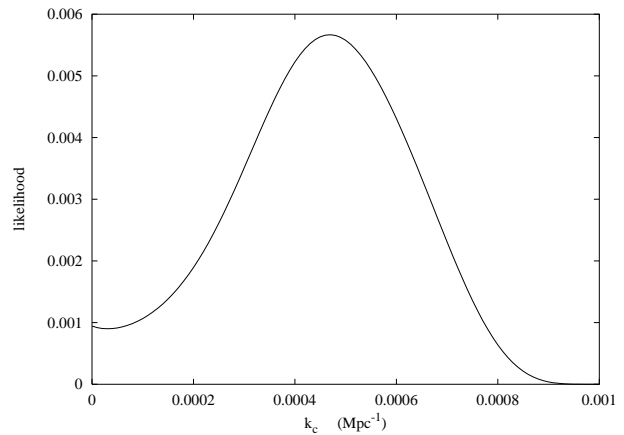}
  \caption{The likelihood as a function of the cutoff wavenumber $k_c$
    for the model of Section~\ref{sec:cutoff}.}
  \label{fig:Likecutoff}
\end{figure}

In Figure~\ref{fig:Clcutoff} we show the CMB power spectrum at low
multipoles with several cutoff models, and in Figure~\ref{fig:Likecutoff}
we show the CMB likelihood as a function of the cutoff scale, $k_c$.
These figures essentially reproduce the results of Contaldi et al.

It is clear that the cutoff allows for a better fit than the so-called best fit.
Next we evaluate the evidence for this model with $k_c$ as the
only parameter, with the prior $p(k_c)\equiv P(k_c|{\rm cutoff})$ chosen to be flat in
the region $[0,0.001]\;\mbox{Mpc}^{-1}$. We obtain
\begin{equation}
  P(D|{\rm cutoff})=\int d k_c \; p(k_c) \;
  {\cal L}(k_c)=0.0025 \ .
\end{equation}
This value is $2.6$ times the evidence for the fiducial model, which implies
that the cutoff model is preferred only at approximately $1.4\sigma$ level.
We have also calculated the Ockham factor for this model, defined in
Eq.~(\ref{eq:Ockham}), to be $0.441$.

Choosing a flat prior over this region emphasizes values of the cutoff
near $k_c\sim0.5\times 10^{-3}\;\mbox{Mpc}^{-1}$, so in fact implements
a sort of fine tuning. We might instead use a prior proportional to
$1/k_c$ (i.e., linear in $\ln k_c$), which also has the advantage of
having the same form if we switch variables to the cutoff length
$l_c\propto1/k_c$. If we choose a lower limit of
$10^{-4}\;\mbox{Mpc}^{-1}$, the evidence is unchanged from the value for
the flat prior, but as we decrease the lower limit the evidence becomes
dominated by the plateau at $k_c\to0$, where the models approach the
fiducial best fit. The limiting value of the evidence is thus the same
value as for the fiducial model itself: the maximum likelihood for this
model may be quite large, but the Ockham factor is small.

\subsection{Geometry: A Closed Universe}
\label{sec:closed}

CMB measurements indicate that the geometry of the universe is
very nearly flat. This is consistent with the inflationary
paradigm in which the universe, unless additionally fine-tuned,
would be expected to be infinitesimally close to flat today.
However, a slightly closed universe is also consistent with the
current data and is actually marginally preferred by the WMAP
experiment \cite{WMAP03Spergel}, whose best fit value was
$\Omega_k=-0.02\pm 0.02$.

When calculating theoretical predictions for CMB anisotropy
spectra one is faced with the so-called geometric degeneracy
among the values of matter density, curvature and dark energy density
\cite{EfstBondConfusion99}. Given fixed values for $\Omega_{\rm cdm}h^2$,
$\Omega_{b}h^2$ and acoustic peak location parameter one can produce almost identical CMB spectra by
choosing the values of $h$ and $\Omega_k$ along a degeneracy line
in the $(h,\Omega_k)$ space. The differences between spectra are
only notable on large scales ($\ell\lesssim 20$) where the ISW
contribution to the anisotropy due to the dark energy component is
dominant.

\begin{figure}[tbp]
  \centering\includegraphics[width=0.67\columnwidth,angle=270]{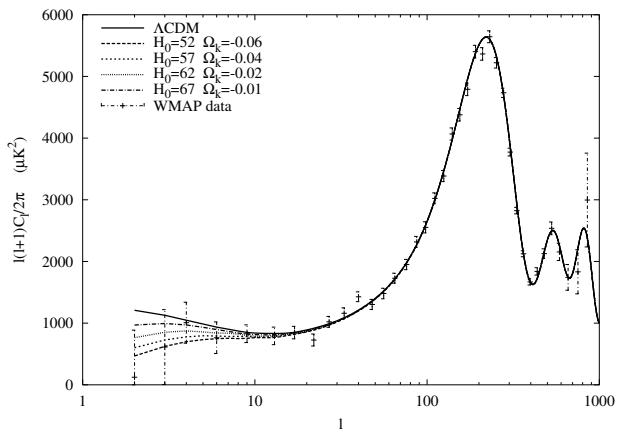}
\caption{The CMB power spectrum for different curvature values in the
  closed model of \ref{sec:closed}.}
  \label{fig:closedcl}
\end{figure}
\begin{figure}[tbp]
  \centering\includegraphics[width=0.67\columnwidth,angle=270]{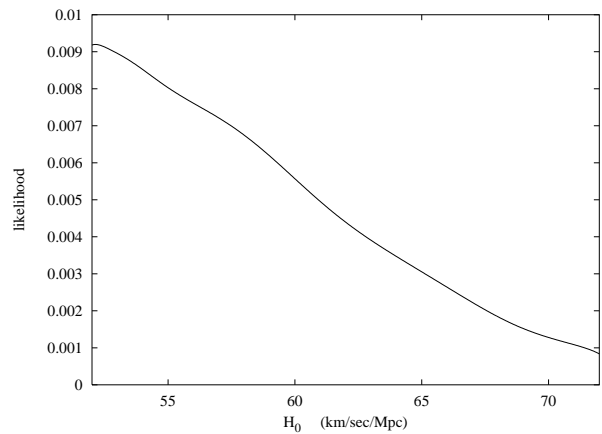}
  \caption{The likelihood as a function of $H_0$ for the
    closed model of \ref{sec:closed}.}
  \label{fig:closed_like}
\end{figure}

A closed universe contains a characteristic scale - the curvature
scale $R_c$. The eigenvalues $\beta$ of the Laplacian are,
therefore, discrete and related to the physical wave-number $k$
via $\beta^2 = 1+ k^2 R^2$ with modes corresponding to $\beta=1$
and $2$ being unphysical pure gauge modes. As argued in
\cite{Efstathiou_lowell_03a}, if the universe was indeed marginally closed, in
the absence of a concrete model it is not obvious how the concept
of scale invariance should be extended to scales comparable to the
curvature scale. One of the possibilities could be that the
spectrum would truncate on scales close to $R$. A heuristic
formula for the primordial spectrum, illustrating such a
possibility, was suggested in \cite{Efstathiou_lowell_03a}:
\begin{equation}
  \label{eq:closedps} P(\beta) \propto {(\beta^2-4)^2 \over {\beta
      (\beta^2-1)}} \left[ 1-{\rm exp} \left( - {{\beta-3}\over 4}
    \right) \right] \ .
\end{equation}
We have used Eq.~(\ref{eq:closedps}) to
generate CMB anisotropy spectra for various values of $\Omega_k$
chosen to lie along the same geometrical degeneracy line that
contained WMAP's best fit flat $\Lambda$CDM model. The results are
shown in Figure~\ref{fig:closedcl}.

As can be seen from the plot, the truncated closed models
fit the data considerably better than WMAP's best fit
model.

Next we calculate the evidence for this model with $h$ as the free
parameter. The prior $p(h)$ was taken to be a Gaussian with the mean
$\bar{h}=0.72$ and variance $\sigma_h=0.10$, and additionally
constrained to be in the range $[0.52,0.72]$.  The lower bound is
dictated by current experimental constraints on the value of $h$, while
the upper bound follows from the fact that along the geometric
degeneracy line higher values of $h$ would correspond to $\Omega_k \ge
0$. We find that the evidence for this model is
\begin{equation}
  P(D|{\rm closed})=\int d h \; p(h) \; {\cal L}(h)=0.0034 \ ,
\end{equation}
where ${\cal L}(h)$ is the likelihood of data given a particular value of $h$.
The obtained  evidence is approximately 3.6 times that of
WMAP's best fit model. This can be interpreted as the closed model being
preferred over the best fit model at a $1.6\sigma$
level, which, considering the absence of a robust model of a marginally closed
universe, is insufficient to warrant abandoning simple
inflation as the base model for fitting data.
The Ockham factor for this model (Eq.~(\ref{eq:Ockham})) is $0.370$.
\begin{figure}[tbp]
  \centering\includegraphics[width=0.67\columnwidth,angle=270]{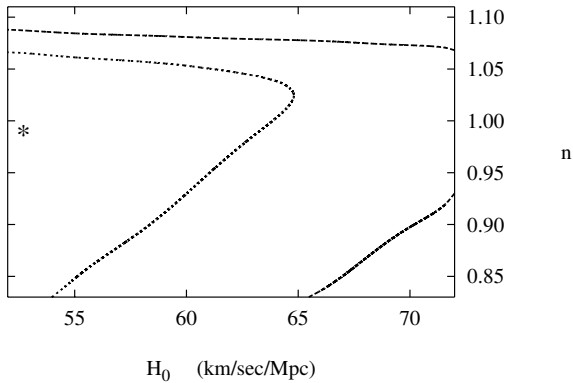}
  \caption{Likelihood contours in the $(n,h)$ parameter space for the
    closed model of Section~\ref{sec:closed}, marginalized over the
    value of $\sigma_8$.  Shown are the 1 and 2$\sigma$ contours,
    defined by the equivalent likelihood ratio for a two-parameter
    Gaussian distribution. The point that maximizes the likelihood
    function is marked with an asterisk ({\large *}).}
  \label{fig:closedlike}
\end{figure}

In addition, we have considered the same closed universe model but with
the spectral index $n_s$ and the value of $\sigma_8$ also allowed to
vary, to see if the fit could be improved further. The prior on $n_s$
was chosen to be Gaussian with $\bar{n}=0.97$ and $\sigma_n=0.07$ and
restricted to the interval $[0.83,1.11]$. The prior on $\sigma_8$ was
also Gaussian with the mean value of $0.95$ and variance $0.05$
restricted to the range $[0.6,1]$. We found the evidence in this case to
be
\begin{eqnarray}
\nonumber
P(D|{\rm closed})&=&
\int d n \; d h \; d \sigma_8 \; p(n) \; p(h) \; p(\sigma_8) \; {\cal L}(n,h,\sigma_8) \\ &=&0.0008 \ ,
\end{eqnarray}
which is lower than the evidence for the fiducial model. The likelihood contours for
this model, after marginalizing over $\sigma_8$, are shown
in Figure \ref{fig:closedlike}. This illustrates how adding more parameter
freedom can dramatically dilute the evidence for the model, even if it
fits the data very well. This is reflected in a very low value of the
Ockham factor for this model, which is only $0.069$.

\section{Theoretical and Experimental Systematics}
\label{sec:systematics}

Having examined the possibility that the observed lack of power on large
scales points in the direction of new physics, we now turn to the
alternative that it can be attributed to data analysis methodology. The
simplest case would be an underestimation of the errors
corresponding to low multipoles. This would mean that we live in a
universe described by the best fit power law model and that the
discrepancy between its predictions and the WMAP data emanates from our
miscalculating the aforementioned errors. Of course, we have copious
evidence from the work done by the WMAP team itself as well as from
comparison with other data that their data is likely to be reliable on
these scales.  Conversely, we could instead interpret this as saying
that the $\ell=2,3$ multipoles are correctly measured, but have an
unknown origin outside the standard cosmology. That is, there is some
model like those considered in the previous sections, but we do not yet
know what it is.

\begin{figure}[bp]
  \centering\includegraphics[width=0.67\columnwidth,angle=270]{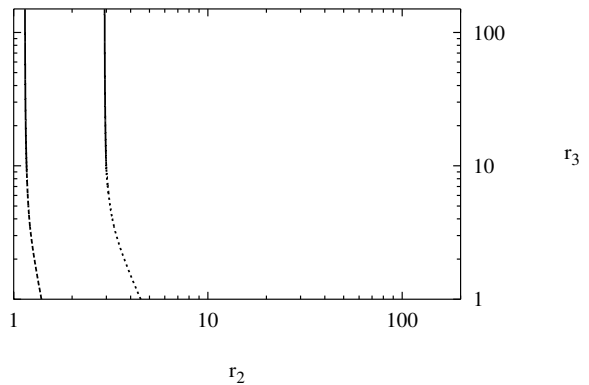}
  \caption{Contours of the likelihood as a function of the parameters
    $r_2$ and $r_3$. Shown are the 1 and 2 $\sigma$ contours. The
    likelihood is maximized in the upper right corner, where $r_2$ and
    $r_3$ are largest.}
  \label{fig:largeslike}
\end{figure}

We implement this idea by multiplying the diagonal elements of the
curvature matrix corresponding to $C_2$ and $C_3$ by two constants
(hereafter referred to as $r_2$ and $r_3$) that serve as the free
parameters of our model. This has the effect of increasing the error bars
of $C_2$ and $C_3$. Figure \ref{fig:largeslike} shows contours of the
likelihood function for various values of these parameters.  We have
also evaluated the evidence for this model to be
\begin{equation}
  P(D|{\rm syst.}) = \int d r_2 \; d r_3 \; p(r_2) \; p(r_3) \;
  {\cal L}(r_2,r_3)=0.0387 \ ,
\end{equation}
using flat priors on $r_2$ and $r_3$ in the
intervals [1,200] and [1,150] respectively; these maxima are chosen for
numerical convenience but the results are insensitive to them as long as
$r_i\gg1$. It is also insensitive to whether we use a uniform prior on
the $r_i$ or on $\ln{r_i}$. The latter are equivalent to
$P(r_i)\propto1/r_i$, the so-called ``Jefferys prior'' appropriate for a
scale parameter.

Note that the likelihood is maximized when these parameters reach their
largest values: the data always become more likely when the error bars
increase. Indeed, this implies that we can consider an even simpler
model with parameters fixed at $r_i\to\infty$. This model has a
likelihood of $0.0414$, giving it a Bayes factor of 44 with respect to
the conventional best fit.  This model corresponds to ignoring the data
at $\ell=2,3$: there is \emph{no model} that can improve the fit here by
more than this roughly $2.75\sigma$ level. 

The evidence for these models implies that
if the correct model at low $\ell$ was indeed other than the ``best
fit'', there would be a roughly $2.75\sigma$ level evidence that the
error bars on $C_2$ and $C_3$ were underestimated.

\section{Discussion}
\label{sec:discussion}

\begin{table}[htbp]
\vskip 0.5 truecm
\begin{tabular}{|c||c|c|c|} \hline
{\bf Model} \  & \ {\bf Ockham factor} \ & \ {\bf Bayes factor} \
& \ {\bf $\sigma$} \ \\ \hline\hline Best fit \ & \ -\ & \ 1 \ & \
- \ \\ \hline Flat with cutoff \ & \ 0.441 \ & \ 2.66 \ & \ 1.40
$\sigma$ \ \\ \hline Closed ($h$) \ & \ 0.370 \ & \ 3.62 \ & \
1.60 $\sigma$ \ \\ \hline Closed ($h$ $\sigma_8$ $n$) \ & \ 0.069
\ & \ 0.85 \ & \ 0.57 $\sigma$ \ \\ \hline Large error bars \ & \
0.945 \ & \ 41.2 \ & \ 2.73 $\sigma$ \ \\ \hline
\end{tabular}
\caption{Summary of the results of the paper. The Bayes factors, $B$, are all
  defined with respect to the ``Best fit'' model of the first row, and
  the column ``$\sigma$'' is defined as $\sqrt{2\left|\ln B\right|}$. The Ockham factors
  are defined in the text, Section~\ref{sec:intro}.}
\label{table}
\end{table}

We summarize our results in Table \ref{table}, presenting the
Bayes and Ockham factors for the models we have discussed. Note
that these numbers explicitly do not consider prior information
about these models. Indeed, all of these models were explicitly
constructed in response to the observed low power. In particular,
the models with low primordial power considered in
Section~\ref{sec:lowpower} require that the scale of the power
cutoff be fine tuned with respect to the horizon scale in order to
reduce power at just the right angular scale, either by fiat or by
determining the location of the curvature scale. {\sl A priori},
such models would be strongly disfavored. However, it has been
recently pointed out in Ref.~\cite{Kesden03}, that a
cross-correlation between CMB and cosmic-shear patterns, as well
as between CMB and low-redshift tracers of the mass distribution,
can provide a supplemental evidence for a large scale cutoff in
the primordial spectrum. Such a cutoff would generally increase
the cross-correlation.

There are models with similar characteristics that have been
discussed separately from these low-power issues: the class of
models with non-trivial topologies
\cite{sokolov93,stevens93,JingFang94,doc95,graca02,LevinPhysRpt,deOli_WMAP03}.
We might assign a greater prior to such models, although again to
explain the observations requires fine tuning of the topology
scale.  In a recent paper de Oliveira-Costa et al
\cite{TegOliHam_WMAP03} argued that the low power on large scales
is unlikely to be a sign of non-trivial topology. We did not
include these models into our analysis; however, one can expect
them to have a similar evidence to the cutoff models we have
considered. Indeed, the type of CMB spectra that these two models
produced are essentially the same and the difference in the values
of the evidence comes mainly from the prior on the free parameter.
Note that models with non-trivial topology will also have other
signatures, possibly observable in the CMB by considering
properties beyond the power spectrum (see e.g.,
\cite{LevinPhysRpt} and references therein).

Other analyses of these data have reached similar conclusions. In
Ref.~\cite{gaztanaga03} Gazta\~naga {\it et al} performed a full
covariance analysis of the WMAP data using the 2-point angular
correlation and its higher-order moments.  They have argued that the
WMAP data is in a reasonable agreement with the $\Lambda$CDM model if
WMAP data was considered as a particular realization of realistic
$\Lambda$CDM simulations with the corresponding covariance.

We have also considered a model that considers a possible systematic
error in the determination of the large-scale power. Although this model
is experimentally unlikely, we can instead consider it as the
\emph{reductio ad absurdum} of all the possibilities we are considering:
what happens if we just throw away the large scale data? From the Bayes
factor of about 44 we see that there is likely \emph{no model} at all
that will ever improve the fit to the large scale by more than about
$2.75\sigma$, in agreement with the somewhat different analysis of
\cite{Efstathiou_lowell_03b}, and to some extent with that of the WMAP
team itself \cite{BennettWMAPbasic03,WMAP03Spergel}.  It is worth noting
that the phases of low harmonics could provide additional information
about the plausibility of a cosmological model; for instance, a model
predicting an alignment of the $\ell$=2,3 harmonics (according to
\cite{deOli_WMAP03}) would be favoured with respect to a model making no
such prediction, given that both models had the same power at low
$\ell$. But we should point out that features like the alignment of the
low harmonics would not have any impact on the power at large scales.
Consequently, no model will ever fare better than about $2.75\sigma$ as
far as power at large scales is concerned.

However, there are other possibilities for probing the physics on the
largest scales. In particular, a better measurement of the polarization
of the CMB and its correlation with the intensity at these same
multipoles will certainly enable us to cement the interpretation of the
anisotropy at the same scales.

\acknowledgments We thank the referee for helpful comments. AHJ would
like to thank G.~Efstathiou for helpful conversations. AHJ and LP were
supported by PPARC in the UK, and AN by EC network HPRN-CT-2000-00124,
``CMBnet''.

\newcommand{\mnras}{Mon.\ Not.\ R.\ Astr.\ Soc.}
\newcommand{\physrep}{Physics Reports}
\bibliography{cmb}

\end{document}